\documentclass[11pt]{article}
\usepackage{amsmath}
\usepackage{amsfonts}
\usepackage{graphicx}
\usepackage[sort,compress]{cite}

\begin{document}
\begin{center}
\Large \bf{Implicit particle methods and their connection with variational data assimilation} 
\end{center}

\begin{center}
Ethan Atkins$^{1}$, Matthias Morzfeld$^2$ and Alexandre J. Chorin$^{1,2}$\\
\vspace{3mm}
Department of Mathematics\\
$^1$University of California, Berkeley\\
and\\
$^2$Lawrence Berkeley National Laboratory
\end{center}
\vspace{1mm}

\begin{center}
\emph{Abstract}
\end{center}
The implicit particle filter is a sequential Monte Carlo method for data assimilation that guides the particles to the high-probability regions via a sequence of steps that includes minimizations. We present a new and more general derivation of this approach and extend the method to particle smoothing as well as to data assimilation for perfect models. We show that the minimizations required by implicit particle methods are similar to the ones one encounters in variational data assimilation and explore the connection of implicit particle methods with variational data assimilation. In particular, we argue that existing variational codes can be converted into implicit particle methods at a low cost, often yielding better estimates, that are also equipped with quantitative measures of the uncertainty. A detailed example is presented.

\section{Introduction}
The goal in data assimilation is to estimate the state of a system by combining information from incomplete and noisy observations of this state with information from a possibly uncertain numerical model. This can be done by analyzing the conditional probability density function (pdf) for the state given the observations \cite{Doucet2001,Kalnay,EvensenBook,ChorinHald}. If the model is linear and the observations are linear functions of the state and if, in addition, all error statistics are Gaussian, then the state conditioned on the data is also Gaussian. In this case, all one needs to know is the mean and covariance of the state and both can be computed by the Kalman filter \cite{Kalman1960,Kalman1961}. However, many problems are nonlinear and non-Gaussian, and methods that assume a nearly linear model or nearly Gaussian errors, such as the ensemble Kalman filters \cite{EvensenBook, EvensenEnKF}, can perform poorly if these assumptions are violated \cite{MillerTellus}. 

For that reason we focus on variational data assimilation and particle methods, which do not require Gaussianity or linearity approximations. In variational data assimilation one finds the most likely state given the observations, i.e. the mode of the conditional pdf, through minimization of a suitable cost function \cite{TalagrandCourtier,DimetTalagrand,tremolet,Bennet1993}. While there is no guarantee that the most likely state is found (the minimization may not converge to the global minimum), variational methods have proven effective in many applications and they are widely used in geophysical data assimilation, e.g. in numerical weather prediction.

Particle methods assimilate the data via Monte Carlo importance sampling \cite{Doucet2001,GordonReview,GordonSIR}. Most particle methods first sample a given importance function and then use the data to assign weights to each sample, so that the weighted samples, called particles in this context, form an empirical estimate of the conditional pdf. The difficulty is that the importance function and the conditional pdf can become nearly mutually singular as the dimension of the state space becomes large, which leads to a representation of the conditional pdf by a single and often uninformative particle \cite{BickelBootstrap,Bickel}. This effect is known as sample impoverishment and has been an obstacle to the application of particle methods to geophysical data assimilation, where the state dimension is typically large. 

Sample impoverishment can be delayed or even prevented if the overlap between the importance function and the conditional density is increased, and much effort has been invested to find an importance function that can work in large dimensional problems, particularly in geophysical applications \cite{Doucet,JohansenDoucet,vanLeeuwen,PeterJan2011,Carpenter}. The implicit particle filter \cite{chorintu2010,chorintupnas,Morzfeld2011} attempts to prevent sample impoverishment by focussing the particles to regions of high probability. These regions are identified through particle-by-particle minimizations. Since the minimization for each particle of an implicit particle filter is similar to the minimizations one encounters in variational data assimilation, one can expect a link between these two approaches. We will describe this link in this paper.

The paper is structured as follows. In section~\ref{sec:implicit sampling}, we review how to sample a given pdf using implicit sampling by first finding the mode of the pdf and then generating samples in the neighborhood of this mode. In section~\ref{sec:implicit filter}, we apply implicit sampling to the conditional pdf for data assimilation to derive the implicit particle smoother that assimilates all available observations at the same time, and the implicit particle filter that assimilates the data sequentially. In section \ref{sec:var comp}, we make the connection between these implicit particle methods and variational data assimilation, and show how existing variational codes can be used for the efficient implementation of implicit particle methods. In section~\ref{sec:results} we present an application of implicit particle methods and discuss their variational aspects. Conclusions are offered in section \ref{sec:Conclusions}.

\section{Implicit sampling}
\label{sec:implicit sampling}
Importance sampling is a Monte Carlo method that samples a hard-to-sample pdf $p$ using an easy-to-sample pdf $p_0$ \cite{Hammersley,KalosWhitlock,ChorinHald,Doucet2001,Geweke}. In this context, the density $p$ we want to sample, but cannot sample directly, is called the target density and the density $p_0$ we actually use to obtain a sample is called the importance density (or importance function). Suppose we are interested in the pdf $p$ of a $d$-dimensional, continuous random variable $x$. Importance sampling generates samples $X\in R^d$ (we use capital letters to denote realizations of random variables) from the known importance function $p_{0}$ with weights 
\begin{equation}
\label{eq:ImportanceSamplingWeights}
	w(X) = \frac{p(X)}{p_0(X)},
\end{equation}
such that the weighted samples form an empirical estimate of the target pdf~$p$. This empirical estimate approximates the target pdf weakly, i.e.
\begin{equation}
\label{eq:ImportanceSamplingMean}
\hat{E}_M(u) = \frac{\sum_{j=0}^{M}u(X_j)w(X_j)}{\sum_{j=0}^{M}w(X_j)},
\end{equation}
converges almost surely to the expected value, $E_p\left[u(x)\right]$, of a function $u$ with respect to the density $p$, as the number of samples, $M$, approaches infinity. It should be clear that the support of $p_0$ must include the support of $p$.  Moreover, importance sampling works even if the target pdf is known only up to a multiplicative constant, because this constant can be eliminated by scaling the weights so that their sum equals one. 

The efficiency of importance sampling depends on the choice of the importance function. For example, samples with a small weight contribute very little to the approximation of the expected value in (\ref{eq:ImportanceSamplingMean}), so that the computational effort spent on generating these low probability samples is mostly wasted. To avoid spending computation time on low probability samples, one needs to find an importance function $p_0$ such that the variance of the weights in (\ref{eq:ImportanceSamplingWeights}) is small, i.e. all samples contribute equally to the sum in~(\ref{eq:ImportanceSamplingMean}). This means in particular that the importance function must be large in the regions where the target density is large. Implicit sampling is an importance sampling method that defines the importance function implicitly by an algebraic equation. We will now show that this importance function is large where $p$ is large, i.e. that the samples we obtain have a high probability.

We write the pdf we are interested as $p\propto e^{-F(x)}$ (this is natural in data assimilation, see section \ref{sec:Problem formulation}) and, for a moment, assume that 
\begin{equation}
\label{eq:DefF}
	F(x)=-\log p(x),
\end{equation}
is convex (we will relax this assumption later on). The region where $p$ is large, and where the high-probability samples lie, is the neighborhood of the mode of $p$. Using the log-transformation (\ref{eq:DefF}), we can identify this region through minimization of $F$, and define
\begin{equation}
	\phi_F = \min F. \nonumber
\end{equation}
To obtain a sample in the high-probability region, we pick a reference variable $\xi \sim g$, with a known pdf $g\propto e^{-G(\xi)}$, and which is easy to sample. We then map the high probability region of the reference variable $\xi$ to the high probability region of $X$. This can be done by solving the algebraic equation
\begin{equation}
	\label{eq:ImplicitEquationGeneral}
	F(X)-\phi_F=G(\xi)-\phi_{G},
\end{equation}
where  $G = -\log g$ is chosen to be convex and $\phi_G=\min G$. Note that the above scalar equation is underdetermined (it connects the $d$ elements of $x$ to the $d$ elements $\xi$) and solvable since $F$ and $G$ are infinite at $\pm \infty$, so that the left and right hand sides of (\ref{eq:ImplicitEquationGeneral}) both range from $\left[0,\infty \right)$. We can thus find a sample $X$ by solving (\ref{eq:ImplicitEquationGeneral}) with a one-to-one and onto mapping
\begin{equation}
\label{eq:Map}
 \psi : \xi  \rightarrow X.
\end{equation}
A sample of the reference density $\xi$ is likely to lie near the mode of $g$, so that the right hand side of (\ref{eq:ImplicitEquationGeneral}) is likely to be small. Equation (\ref{eq:ImplicitEquationGeneral}) and the mapping $\psi$ thus imply that, for a high-probability sample of $\xi$, the function $F(X)$ is close to its minimum $\phi$, which implies that $X$ is in the region where $p$ is large. The map $\psi$ thus maps the high-probability region of the reference variable $\xi$ to the high-probability region of $X$, so that, with a high probability, we obtain a high probability sample.

The reference variable $\xi$ and the map $\psi$ in (\ref{eq:Map}) define the importance function
\begin{equation}
	p_0(X(\xi))  \propto \frac{\exp(-G(\xi))}{  \left| J(\xi) \right|}, \nonumber
\end{equation}
where $J=\det (\partial X/\partial \xi )$ is the Jacobian of $\psi$. Using~(\ref{eq:ImplicitEquationGeneral}), the importance function can be written in terms of $X=\psi(\xi)$
\begin{equation}
	\label{eq:ImportanceFunction}
	p_0(X) \propto \frac{\exp(-F(X)+\phi_F-\phi_G)}{\left|J(X)\right|},
\end{equation}
and, by using (\ref{eq:ImportanceSamplingWeights}), we find that the weight of the sample $X$ is 
\begin{equation}
\label{eq:weightsGeneral}
	w(X) \propto   e^{-\phi_F+\phi_G} \left|J(X)\right|.
\end{equation}
The variability in the weights is induced by the Jacobian of the map (the term involving the $\phi$'s is constant among the samples and can be removed by scaling the weights so that their sum equals one). The only requirement on $\psi$ is that it solves the undetermined equation (\ref{eq:ImplicitEquationGeneral}). We thus have a lot of freedom in choosing this map and we can use this freedom to construct a map that keeps the variance of the weights small, and whose Jacobian is easy to compute. Various ways of doing this have been presented in \cite{chorintu2010,chorintupnas,Morzfeld2011} and we will review one of these maps in the example in section~\ref{sec:results}. What is important here is to realize that solving the algebraic equation (\ref{eq:ImplicitEquationGeneral}) is cheap, once the minimum of $F$ is found.

What is left to do is to choose a reference variable $\xi$. Equation (\ref{eq:ImportanceFunction}) implies that the closer the pdf of the reference variable resembles the target density~$p$, the more the importance function $p_0$ also resembles the target density. It is thus desirable to choose such a reference variable, however that might be impractical (because we typically do not know the target pdf in advance and because it is hard to sample the target pdf). In practice one should choose a reference density that is easy to sample and easy to minimize. For example, in \cite{chorintu2010,chorintupnas,Morzfeld2011}, a Gaussian reference variable, $\xi\sim\mathcal{N}(0,I)$, was used and yielded good results (we denote a Gaussian variable with mean $\mu$ and covariance matrix $\Sigma$ by $\mathcal{N}(\mu,\Sigma)$ and use $I$ for the identity matrix of appropriate dimensions). A Gaussian reference variable does not imply that the target density is approximated by a Gaussian, since it is clear from (\ref{eq:ImportanceFunction}) that the importance density is generally not Gaussian even if $\xi$ is. Instead, each sample $X$ is a function of a Gaussian reference sample.

We now relax the assumption that $F$ is convex. If $F$ is $U$-shaped, then the above construction works without modification. A scalar function $F$ is called $U$-shaped if it is at least piecewise differentiable, its first derivative vanishes at a single point which is a minimum, $F$ is strictly decreasing on one side of the minimum and strictly increasing on the other, and $F(X)\rightarrow \infty$ as $\left|X\right| \rightarrow \infty$; in the $d$-dimensional case, $F$ is $U$-shaped if it has a single minimum and each intersection of the graph of the function $y = F(X)$ with a vertical plane through the minimum is $U$-shaped in the scalar sense. If $F$ is not $U$-shaped, but has only one minimum, one can replace it by a $U$-shaped approximation, say $F_0$, and then apply implicit sampling as above. The error one makes by this approximation can be accounted for through reweighting \cite{chorintu2010}. If $F$ has multiple minima (the target pdf $p$ has more than one mode), then one can find local $U$-shaped approximations at each local minimum and apply implicit sampling to each local approximation. The errors one makes can be accounted for by reweighting of the samples.

\section{Implicit sampling for data assimilation}
\label{sec:implicit filter}
We now apply implicit sampling to the conditional pdf for data assimilation and derive three implicit particle methods. Our derivation is more general than the ones presented in \cite{chorintupnas,chorintu2010,Morzfeld2011} and highlights the variational aspects of the implicit particle methods.

\subsection{Problem formulation}
\label{sec:Problem formulation}
We start with a review of the data assimilation problem to set up notation and terminology. In data assimilation, one is given an uncertain numerical model of a system and a stream of noisy data about its state, and one wants to use both to estimate the state of the system. The numerical model is a Markovian state space model 
\begin{equation}
\label{eq:DiscreteModel}
x_{j+1}=R_j(x_{j}) + G_j(x_{j})Z_{j},
\end{equation}
where $j=0,1,2,\dots$ can be thought of as discrete time; the state, $x_j$, is a $d$-dimensional real vector, $R_j$ respectively $G_j$ are given $d$-dimensional vector functions, respectively real $d\times d$ matrices and the $Z_j$'s are $d$-dimensional random variables. In geophysical applications, the numerical model often comes from discretizations of stochastic differential equations, in which case the $Z_j$'s are random vectors whose elements are independent normal variates~\cite{Kloeden}, and we assume the $Z_j$'s to be Gaussian from now on. We assume further that at time $j=0$ the pdf for the state $x_0$ is known and that the matrices~$G_j$ have full-rank. How to relax the latter assumption is described in \cite{Morzfeld2012}.

The data 
\begin{equation}
\label{eq:obs}
y_{k}=h(x_{n_{k}}) +V_k,
\end{equation}
indexed by $k=1,2,\dots$, are regularly spaced, noisy measurements of the state, taken at times $n_k = kr$, where $r\geq 1$ is a positive integer (it is an easy exercise to consider also the case when observations are irregularly spaced in time). In the above equation, $h$ is a $b$-dimensional vector function and $V_k$ is a $b$-dimensional random variable with a known pdf.  We assume that the random variables $V_k$ are independent of each other and also independent of the model noise $Z_j$. For notational convenience, we will write $x_{0:k}$ for the sequence of vectors $x_0,\dots,x_k$.

At time $n_m=n\cdot m$, $m\geq1$, we have collected $m$ observations $y_{1:m}$, and everything we know about the state trajectory $x_{0:n_{m}}$ is contained in the conditional pdf\begin{equation}
p(x_{0:n_{m}}|y_{1:m})=p(x_0)\frac{\prod_{j=1}^{n_m}p(x_j|x_{j-1})\prod_{j=1}^{m}p(y_j|x_{n_j})}{p(y_{1:m})}. \label{eq:CondPDF}
\end{equation}
Since we know $p(x_0)$, and can read $p(x_j|x_{j-1})$ and $p(y_j|x_j)$ from equations~(\ref{eq:DiscreteModel}) and (\ref{eq:obs}), we know this pdf up to the normalization constant $p(y_{1:m})$.

\subsection{The implicit particle smoother} 
\label{sec:smoother}
To assimilate all $m$ available observations, we can apply implicit sampling to the conditional pdf in (\ref{eq:CondPDF}). Since an importance sampling scheme that assimilates more than one observation at a time is often called a particle smoother \cite{Doucet2001}, we will call this method the implicit particle smoother.

The target pdf is the conditional pdf in (\ref{eq:CondPDF}), so that the function $F$ of implicit sampling is 
\begin{equation}
F(x_{0:n_m})= -\log(p(x_{0:n_m}|y_{1:m})). \nonumber
\end{equation}
If $V_k$ in (\ref{eq:obs}) is Gaussian with mean zero and  covariance matrix $Q$, then this $F$ is
\begin{align}
F(x_{0:n_m}) & = -\log(p(x_0))\nonumber \\
 &\quad +\frac{1}{2}\sum_{j=0}^{n_{m}-1}(x_{j+1}-R_{j}(x_{j}))^T\Sigma_{j}^{-1}(x_{j+1}-R_{j}(x_{j})) \nonumber \\
 &\quad +\frac{1}{2}\sum_{j=1}^{m}(y_j-h(x_{n_{j}}))^{T}Q^{-1}(y_j-h(x_{n_{j}}))+C,\label{eq:SmootherF}
\end{align}
where $\Sigma_j = G_j(x_j)^TG_j(x_j)$, and where the value of the constant $C$ is irrelevant (it will drop out in the normalization of the weights). We find the minimum $\phi_F$ of $F$ using standard techniques, such as Newton's methods, quasi Newton methods or gradient descent (see e.g. \cite{Conn,Fletcher,Nocedal}) and choose a Gaussian reference variable $\xi\sim \mathcal{N}(0,I)$. In this case the algebraic equation~(\ref{eq:ImplicitEquationGeneral}) becomes
\begin{equation}
F(X)-\phi_F= \frac{1}{2}\xi^T\xi,
\label{eq:ImplicitEquation}
\end{equation}
which we solve with a suitable mapping $\psi$ (see \cite{chorintu2010,chorintupnas,Morzfeld2011}) for $M$ independent realizations of $\xi$ to obtain $M$ weighted samples (particles), with weights given by~(\ref{eq:weightsGeneral}). The $M$ particles form an empirical estimate of the conditional pdf $p(x_{0:n_m})$, so that we have successfully assimilated the data. To compute a concrete state estimate, we can compute, for example, the weighted sample average to approximate the conditional  mean $E(x_{0:n_m}|y_{1:m})$, which, in turn, is the minimum mean squared error estimate \cite{ChorinHald}.

\subsection{The implicit particle filter}
\label{sec:filter}
Suppose we have assimilated $m$ observations, for example by using the implicit particle smoother, and that a new observation $y_{m+1}$ is now available. One can of course assimilate this observation by redoing the calculations of the previous section with $p(x_{0:n_{m+1}}|y_{1:m+1})$ replacing  $p(x_{0:n_{m}}|y_{1:m})$, however this approach becomes impractical as we collect more and more data. 

Alternatively, we can assimilate the data sequentially using the recursive formula for the conditional pdf (see \cite{Doucet2001})
\begin{equation}
p(x_{0:n_{m+1}}|y_{1:m+1})=p(x_{0:n_{m}}|y_{1:m})\frac{p(x_{n_{m}+1:n_{m+1}}|x_{n_{m}})p(y_{m+1}|x_{n_{m+1}})}{p(y_{m+1}|y_{1:m})}. \nonumber
\end{equation}
Given a set of $M$ weighted samples $\{X^{k}_{0:n_{m}},w^{k}\}$ (particles), $k=1,\dots,M$, that form an empirical estimate of the conditional pdf $p(x_{0:n_{m}}|y_{1:m})$, the goal is to update each particle to time $n_{m+1}$, by generating a sample $X_{n_{m+1}:n_{m+1}}$ using an importance function $p_0$, and putting 
$$\{X^{k}_{0:n_{m+1}},w^{k}\}= \{(X^{k}_{0:n_{m}},X^{k}_{n_{m}+1:n_{m+1}}),\hat{w}^k\},$$
with updated weights
\begin{equation}
\label{eq:weight_update}
\hat{w}^{k} = w^{k}\frac{p(X^k_{n_m+1:n_{m+1}}|X^{k}_{n_{m}})p(y_{m+1}|X^{k}_{n_{m+1}})}{p_0(X^k_{n_m+1:n_{m+1}})}.
\end{equation}
The assimilation of data using the above sequential importance sampling approach is known as particle filtering (as opposed to the particle smoother, which does not operate sequentially).

For an efficient particle filter, we need to find an importance function $p_0$ that closely resembles the functions $p(X^k_{n_i+1:n_{i+1}}|X^{k}_{n_{i}})p(y_{i+1}|X^{k}_{n_{i+1}})$ for each particle. We can achieve this by applying implicit sampling to each particle, and we will call this approach the implicit particle filter. Thus, we define $M$ functions $F^k$ by
\begin{equation}
\label{eq:FFilter}
F^k(x_{n_m+1:n_{m+1}})=-\log(p(x_{n_m+1:n_{m+1}}|X^{k}_{n_{m}})p(y_{m+1}|x_{n_{m+1}}))
\end{equation}
For Gaussian observation noise, $V_k \sim \mathcal{N}(0,Q)$, these functions $F^k$ become
\begin{align}
F^k(x_{n_m+1:n_{m+1}})&=\frac{1}{2}(x_{n_m+1}-R_{n_m}(X^k_{n_m}))^T\Sigma_{n_m}^{-1}(x_{n_m+1}-R_{n_m}(X^k_{n_m}))
\nonumber \\
& +\frac{1}{2}\sum_{j=n_{m}+1}^{n_{m+1}-1}(x_{j+1}-R_{j}(x_{j}))^T\Sigma_{j}^{-1}(x_{j+1}-R_{j}(x_{j}))  \nonumber \\
& +\frac{1}{2}(y_{m+1}-h(x_{n_{m+1}}))^{T}Q^{-1}(y_{m}-h(x_{n_{m+1}})) +C 
\label{eq:F_spec}
\end{align}
where $C$ is a constant whose value is irrelevant. We find the minima $\phi_k$ of each of these $F_k$'s using standard techniques, such as Newton's method, quasi Newton methods or gradient descent (see e.g. \cite{Conn,Fletcher,Nocedal}). We then pick a Gaussian reference variable $\xi\sim\mathcal{N}(0,I)$ and obtain $M$ samples, $X_{n_m+1:n_{m+1}}^k$, by solving the $M$ equations
\begin{equation}
	F^k(X_{n_m+1:n_{m+1}}^k) - \phi^k = \frac{1}{2} \xi^T\xi, 
	\label{eq:FImplicitFilter}
\end{equation}
with a suitable mapping $\psi$ (see \cite{chorintu2010,chorintupnas,Morzfeld2011}). The update equation for the weights can be obtained by combining (\ref{eq:weightsGeneral}) with (\ref{eq:weight_update}):
\begin{equation}
\label{eq:weights}
\hat{w}^k=w^ke^{-\phi^k}J(X_{n_m+1:n_{m+1}}^k)
\end{equation}
where $J$ is the Jacobian of $\psi$. We append the $M$ samples $X_{n_m+1:n_{m+1}}^k$ to the $M$ particles we already had, and replace their weight with the updated weight from (\ref{eq:weights}). We thus obtain $M$ updated particles that approximate the conditional pdf $p(x_{0:n_{m+1}}|y_{1:m+1})$ at time $n_{m+1}$. As a concrete state estimate, we can use, for example, the weighted sample average to approximate the conditional  mean. Before the next observation is assimilated, the weights can be eliminated by resampling, using one of the algorithms in e.g. \cite{Doucet2001,LiuChen1995,DelMoral2012,GelfandSmithResamp}.

Note that the term $\exp(-\phi^k)$ in (\ref{eq:weights}) induces additional variability into the weights when compared to the implicit particle smoother in section \ref{sec:smoother}, where the variability of the weights is due to only the Jacobian. The additional factor appears here because we apply implicit sampling to $M$ different functions $F^k$ which arise because of the sequential problem formulation (for the implicit particle smoother, we applied implicit sampling to one function $F$). The functions $F^k$ however differ only in the position of each particle, $X^k_{n_m}$, at time $n_m$ (see equations (\ref{eq:FFilter}) and (\ref{eq:F_spec})). For that reason, the minima $\phi^k$ of these functions should not vary too much from particle to particle, so that the variance induced by this term can be expected to be small. The numerical experiments in section \ref{sec:results}, as well as those in \cite{chorintu2010,Morzfeld2011} confirm this statement, however a rigorous analysis of the variance of the weights of the implicit particle filter has not been reported.

\subsection{The implicit particle smoother for perfect models}
\label{sec:PerfectSmoother}
If model errors are small compared to observation errors, one can put 
\begin{equation}
	G_j(x_j)=0,\nonumber
\end{equation}
in (\ref{eq:DiscreteModel}), so that the state trajectory, $x_{1:n_m}$, is a deterministic function of the initial condition $x_0$. This assumption is often called the perfect model assumption and our goal is to find an initial state that is compatible with the available data $y_k$, $k=1,\dots,m$. 

The implicit particle smoother in section \ref{sec:smoother} can be easily adapted to this situation by applying implicit sampling to the conditional pdf $p(x_0|y_{1:m})$. Using Bayes' theorem, the fact that the observations $y_k$ are independent of each other, and that $x_{1:n_m}$ is a deterministic function of $x_0$, we can rewrite this conditional pdf as
\begin{equation}
	p(x_0|y_{1:m}) \propto p(x_0)\displaystyle \prod_{j=1}^{m} p(y_{j}|x_{n_j}),\nonumber
\end{equation}
where the factors $p(y_{j}|x_{n_j})$ are specified by the observation equation~(\ref{eq:obs}). The pdf $p(x_0)$ is called the prior density and is often chosen to be Gaussian. However, the conditional pdf is generally far from being Gaussian, because~$h$ is generally nonlinear and the $x_{n_{j}}$'s are nonlinear functions of $x_{0}$ (see~(\ref{eq:DiscreteModel})).

For implicit sampling of $p(x_0|y_{1:m})$, we define 
\begin{equation}
	F(x_0) = -\log \left(p(x_0|y_{1:m}) \right),\nonumber 
\end{equation}
which for a Gaussian observation noise, $V_k\sim\mathcal{N}(0,Q)$, becomes
\begin{equation}
	 F(x_0) = -\log  \left(p(x_0)\right)+ \sum_{j=1}^{m}(h(x_{n_{j}})-y_{j})^{T}Q^{-1}(h(x_{n_{j}})-y_{j})+C,\label{eq:Fs4DVar}
\end{equation}
 where the value of the constant $C$ is irrelevant. With this $F$, we can find $M$ samples from $p(x_0|y_{1:n_m})$ by first minimizing $F$ and then solving~(\ref{eq:ImplicitEquation}) repeatedly for $M$ realizations of $\xi$. We can solve this scalar equation efficiently using e.g. random maps as in~\cite{Morzfeld2011}, or one of the methods in \cite{chorintu2010}. What is important to realize here is that sampling is fast, once the minimum of $F$ has been found. 

Finally, we want to point out that the above implicit smoothing algorithm above can be modified to assimilate data sequentially, i.e. assimilate $k< m$ observations at a time. We can assimilate the first $k$ observations, $y_{1:k}$, by implicitly sampling $p(x_0|y_{1:k})$ and use the results to construct an empirical approximation of a ``prior'' density for $x_{n_k}$. With that prior, we repeat the same steps to assimilate the next set of observations $y_{k+1:2k}$ by implicitly sampling $p(x_{n_k}|y_{k+1:2k})$ etc. until all available observations are assimilated. Note that the method naturally keeps track of the uncertainty, whereas 4D-Var codes often use ad-hoc approximations to update the covariance matrices \cite{KalnayEnkFvs4DVar}. A sequential approach for data assimilation for perfect models is important in many applications with very large data sets, e.g. in numerical weather prediction or geomagnetics~\cite{Fournier2010}, however the details, as well as numerical tests for sequential implicit sampling for this problem are deferred a future paper.

\section{Connection with variational data assimilation}
\label{sec:var comp}
Variational data assimilation methods find the most likely state trajectory, given the available observations, i.e. the mode of the conditional pdf $p(x_{0:n_m}|y_{1:m})$. We now make the connection between variational methods and the implicit particle filter and smoother, and show how existing codes for variational data assimilation can be used for efficient implementation of these implicit particle methods. We distinguish between weak and strong constraint variational methods.

\subsection{Connection with strong constraint 4D-Var}
\label{sec:strong4DVar}
Strong constraint 4D-Var, see e.g. \cite{DimetTalagrand,Rabier4DVAR,TalagrandCourtier,Talagrand1997,Courtier1997,Courtier1994}, finds the mode of the conditional pdf $p(x_0|y_{1:n_m})$, where $x_0$ is the unknown initial condition of the discrete model (\ref{eq:DiscreteModel}), by minimization of a suitable cost function. Specifically, if the pdf $p(x_0)$, which is often called the prior density, is Gaussian and if the observation noise is also Gaussian, the strong constraint 4D-Var cost function is
\begin{equation}
\mathcal{J}_s(x_{0})=(x_{0}-x_{b})^{T}B^{-1}(x_{0}-x_{b}) + \sum_{j=1}^{m}(h(x_{n_{j}})-y_{j})^{T}Q^{-1}(h(x_{n_{j}})-y_{j}), \label{eq:s4DVarCost}
\end{equation}
where $x_b\in R^d$, called the background state, is the mean of $p(x_0)$ and $B\in R^{d\times d}$ is the covariance matrix of the background state. 

If the observation operator $h$ is linear, the gradient of the cost function $\mathcal{J}_s$ can be found using the adjoint method (see e.g. \cite{TalagrandCourtier}). With this gradient, we can minimize $\mathcal{J}_s$ efficiently using e.g. gradient descent or quasi Newton methods. In the general case ($h$ not linear), one can linearize $h$ along a state trajectory and use this linearization along with the adjoint method to compute an approximate gradient of $\mathcal{J}_s$. The conditions under which a numerical minimization with an approximate gradient converges to the minimum of the cost function $\mathcal{J}_s$ are not well understood. However the method seems to work in many applications. In fact, the use of the adjoint method makes the minimization of $\mathcal{J}_s$ very efficient and, as a result, strong constraint 4D-Var a powerful method for nonlinear data assimilation.

The strong constraint 4D-Var cost function $\mathcal{J}_s$ in (\ref{eq:s4DVarCost}) is identical to $F$  in~(\ref{eq:Fs4DVar}) (up to irrelevant constants), provided we use the same, and not necessarily Gaussian, prior pdf $p_0$. Turning a strong constraint 4D-Var code into an implicit particle smoother (see section \ref{sec:PerfectSmoother}) thus amounts to adding a sampling and weighting step, which in turn amounts to solving the scalar equation (\ref{eq:ImplicitEquation}), or more generally (\ref{eq:ImplicitEquationGeneral}). Efficient methods for executing the sampling and weighting can be found in \cite{chorintu2010,Morzfeld2011}, so that the additional computational cost of implicit particle smoothing is small. The payoff is that the implicit particle smoother approximates the conditional mean and, thus, minimizes the mean square error, whereas 4D-Var computes the conditional mode. The conditional mean is, under wide conditions, a better state estimate than the conditional mode, particularly if the conditional pdf has significant skew. Moreover, the implicit particle smoother naturally produces a quantification of the uncertainty (because it generates an empirical estimate of the conditional pdf), whereas 4D-Var codes provide at best approximate error estimates~\cite{Rabier4DVAR}. 

When the data are sparse in space or time, the conditional pdf often has more than one mode so that the cost function $\mathcal{J}_s$ has multiple minima. Strong constraint 4D-Var will find one of these minima and return it as the state estimate. Important information from the other modes is lost. The implicit particle smoother on the other hand can perform well in multimodal situations (see sections \ref{sec:implicit sampling} and \ref{sec:results}) and, in theory, represents all modes of the conditional pdf by its samples. In practice, there is no guarantee that the implicit particle smoother can sample all modes in all cases (because the numerical minimization may miss local minima), however the representation of a multimodal conditional pdf by the implicit particle smoother through at least some of its modes is superior to the results of a 4D-Var code, that represents the conditional pdf by only one of its modes.

\subsection{Connection with weak constraint 4D-Var}
Weak constraint 4D-Var (see e.g. \cite{Bennet1993,Kalnay, Kurapov2007}) relaxes the perfect model assumption made in strong constraint 4D-Var. There are several ways of doing so \cite{tremolet}, however we only consider here the ``full'' weak 4D-Var problem, i.e. we choose the model state $x_{0:n_{m}}$ as the control vector. The weak constraint 4D-Var method then computes the most likely state trajectory given the available data $y_{1:m}$, i.e. the mode of the conditional pdf $p(x_{0:n_m}|y_{1:m})$. 

The conditional mode is found by minimizing the weak constraint cost function 
\begin{equation}
\label{eq:w4DVarCost}
\mathcal{J}_w(x_{0:n_m}) = -2\log p(x_{0:n_m}|y_{1:m}). \nonumber
\end{equation}
Specifically, for a Gaussian prior density $p(x_0)\sim \mathcal{N}(x_b,B)$, the weak constraint 4D-Var cost function is
\begin{align}
\nonumber
\mathcal{J}_w(x_{0:n_{m}})&=(x_{0}-x_{b})^{T}B^{-1}(x_{0}-x_{b})\nonumber \\
&+\sum_{j=0}^{n_{m}-1}(x_{j+1}-R_{j}(x_{j}))^T\Sigma_{j}^{-1}(x_{j+1}-R_{j}(x_{j})) \nonumber  \\
& +\sum_{j=1}^{m} (y_{jk}-h(x_{n_{j}}))^{T}Q^{-1}(y_{j}-h(x_{n_{j}})). \label{eq:weak4Dcost}
\end{align}
The adjoint method is not directly applicable to finding the gradient of $\mathcal{J}_w$, but related approximate methods can be devised to streamline and accelerate the minimization, see e.g. \cite{Kalnay,Zupanski1997}. Although it is not yet well understood under what conditions a minimization with such approximations converges to a minimum, weak constraint 4D-Var is widely used and is known to produce good state estimates, in particular in numerical weather prediction.

Note that the cost function $\mathcal{J}_w$ in (\ref{eq:weak4Dcost}) equals $F$ in (\ref{eq:SmootherF}), the function that is minimized by the implicit particle smoother of section \ref{sec:smoother} (up to irrelevant constants). We can thus use a weak 4D-Var code for the implementation of an implicit particle smoother to minimize this $F$. Once the minimum is found, we can obtain $M$ samples from the conditional pdf by solving  (\ref{eq:ImplicitEquation}) repeatedly and at a low cost. The additional cost of implementing the implicit particle smoother versus a weak constraint 4D-Var method is thus not large. The implicit particle smoother however has the advantage that it can return the conditional mean as a state estimate, which, under wide conditions, is a better state estimate than the conditional mode (the result of a weak 4D-Var calculation). Moreover, the state estimate of the implicit particle smoother is equipped with a quantitative measure of its uncertainty.

Recall that the implicit particle filter of section \ref{sec:filter} is an efficient \emph{sequential} sampling method for the conditional pdf. The implicit particle filter requires at each assimilation and for each particle, the minimization of the function $F^k$ in~(\ref{eq:FFilter}). These $F^k$'s are parameterized by the previous position of each particle and by the current observation. Moreover, for each particle, $F^k$ is nearly identical to the cost function $\mathcal{J}_w$ of weak constraint 4D-Var in (\ref{eq:weak4Dcost}). The differences are in the treatment of the background state. It is unnecessary to include the background state in the functions $F^k$ because the implicit particle filter samples the prior directly, and without making a Gaussian assumption. Since the implicit particle filter is a sequential method, we set it up in section \ref{sec:filter} to assimilate one observation at a time, so that the arguments of $F^k$ are $x_{n_{m}+1:n_{m+1}}$. We can thus obtain the $F^k$'s from the weak constraint cost function $\mathcal{J}_s$ in (\ref{eq:weak4Dcost}) by removing the background state, turning the \emph{variables} $x_0$ into \emph{parameters} $X^k_{n_m}$ (the position of the $k$th particle at time $n_m$), and running the variational assimilation over one observation only. The particle-by-particle minimizations of $F^k$ for the implicit particle filter can thus be carried out by existing weak constraint 4D-Var codes with only minor modifications. Once the minimum of each $F^k$ is found, the sampling can be carried out efficiently using the methods in \cite{chorintu2010,Morzfeld2011}, so that this extension comes at an acceptable cost. Moreover, the minimization for each particle can be done in parallel so that the runtime of a parallelized implementation of the implicit particle filter is comparable to serial 4D-Var codes.

The main benefits for the implicit particle filter are (\emph{i}) the implicit particle filter tracks the time evolution of the conditional pdf and, thus, can compute the conditional mean, which minimizes the mean square error; (\emph{ii}) the filter naturally produces a quantitative representation of the uncertainty (because it tracks the conditional pdf); and (\emph{iii}) the implicit particle filter handles new observations (in time) naturally, because it is set up as a sequential method. The last point is particularly important when the data sets are large. 

We argued in the previous section that the improvement of strong constraint 4D-Var by the implicit particle smoother is particularly pronounced if the conditional pdf has more than one mode. The arguments presented towards the end of section \ref{sec:strong4DVar} also hold for the weak constraint problem and we expect the implicit particle filter and smoother to perform better than weak constraint 4D-Var in such cases.

\section{Application to the Lorenz attractor}
\label{sec:results}

To illustrate the ideas of the previous sections, we follow \cite{MillerTellus,EvensenLorenz,Chorin2004} and apply the implicit particle filter of section \ref{sec:filter} and the implicit particle smoother of section \ref{sec:PerfectSmoother} to the Lorenz attractor \cite{Lorenz63}. We distinguish between the strong and weak constraint problem.

\subsection{The strong constraint problem}
\label{sec:StrongProblemExample}
The Lorenz attractor is governed the set of ordinary differential equations~(ODE)
\begin{equation}
\frac{dx^1}{dt} =\sigma(x^2-x^1),\quad \frac{dx^2}{dt} =x^1(\rho-x^3)-x^2,\quad \frac{dx^3}{dt} =x^1 x^2-\beta x^3, \label{eq:Lorenz}
\end{equation}
where $\rho=28$, $\sigma=10$, $\beta=8/3$ \cite{Lorenz63}. We discretize these equations using a fourth-order Runge-Kutta scheme with constant time step $\delta = 0.01$. For our purposes, and because we will consider only relatively short integration times, this approximation is sufficiently accurate.

We observe the variables $x^1$ and $x^3$, corrupted by Gaussian noise with mean zero and covariance matrix $Q=2 I_2$ ($I_m$ is the $m\times m$ identity matrix), every $r=20$ model steps, i.e. every 0.2 dimensionless time units. The observation equation (\ref{eq:obs}) thus becomes
\begin{equation}
y_{k}=\left(x^1(t_{n_k}),x^3(t_{n_k})\right)^T+ V_k,\nonumber
\end{equation}
with $V_k\sim \mathcal{N}(0,Q)$. Our goal is to update the prior knowledge about the initial state $x_0$, which we assume to be Gaussian, so that $p_0 \sim \mathcal{N}(x_b,B)$ with $x_b = \left(4.3735, 6.9590, 15.4321 \right)^T$ and $B=0.5I_3$, based upon $4$ observations $y_1,\dots,y_4$. We try to achieve this goal by using the implicit particle smoother of section \ref{sec:PerfectSmoother}. 

Recall that the implicit particle smoother essentially consists of three steps: (\emph{i}) minimize the function $F$ in (\ref{eq:Fs4DVar}); (\emph{ii}) obtain samples from the underlying conditional pdf by solving the algebraic equation (\ref{eq:ImplicitEquation}); and (\emph{iii}) weighing the samples using (\ref{eq:weightsGeneral}). As pointed out in section \ref{sec:strong4DVar}, the first step can be carried out using adjoint codes and that is what we did for this example.

\subsubsection{Variational implementation of the implicit particle smoother}
We constructed the linear tangent adjoint of the continuous time ODE's in~(\ref{eq:Lorenz}) and discretized the adjoint equations using a fourth order Runge-Kutta scheme with time step $\delta=0.01$. We use these adjoint equations to compute the gradient of the function $F$, which in turn is used in a BFGS method (see e.g. \cite{Nocedal,Fletcher}) for the minimization of $F$. To initialize this BFGS method, we ran a few steps of a BFGS method on the ``maximum likelihood'' problem (i.e. we neglect the background term in $F$), in which we could also use the adjoint equations for the gradient computations. The result of the BFGS iteration on the maximum-likelihood problem was used to initialize the BFGS method for the minimization of $F$. We found that this approach is quicker than using the BFGS method on $F$, initialized with the background state $x_b$, because, for our choice of parameters, $F$ seems to have a rather flat region around the background state which is not the minimum. The minimization converged quickly and, on average, took only about $1$ second (using non-optimized Matlab code). We observed occasionally that the minimization was trapped in very flat regions, in which case we re-started the whole process, using a sample from the prior density $p_0$ to initialize the minimization.

\subsubsection{Implementation of the map $\psi$}
Upon minimization of $F$ with the adjoint method, we solve (\ref{eq:ImplicitEquationGeneral}) to obtain samples from the conditional pdf. To solve this underdetermined equation, we need to define a reference variable $\xi$, as well as a map from $\xi$ to $X$. Here, we follow \cite{chorintu2010} and choose a Gaussian reference variable $\xi\sim\mathcal{N}(0,I_3)$ and consider a map $\psi$ that makes use of the quadratic expansion of $F$,
\begin{equation}
F_0(x)= \phi+\frac{1}{2}(x-\mu)^TH(x-\mu), \nonumber
\end{equation}
where $\mu = \mbox{argmin} F$ is the minimizer of $F$ and $H$ is an approximation of its Hessian, evaluated at the minimizer, which is available from the variational minimization using BFGS. To obtain a sample, we then solve the quadratic equation
\begin{equation}
\label{eq:F_0}
F_0(x)- \phi=\frac{1}{2}\xi^T\xi, 
\end{equation}
instead of (\ref{eq:ImplicitEquation}). This can be done efficiently using the Cholesky factor $L$ of~$H$: 
\begin{equation}
	X=\mu+L^{-T}\xi. \label{eq:SampleLorenz}
\end{equation} 
The Jacobian of this map is easily calculated to be the determinant of $L$ (the product of its diagonal entries) and is constant among the particles. We account for the error we made by solving (\ref{eq:F_0}) rather than (\ref{eq:ImplicitEquation}) by reweighing the samples such that
\begin{equation}
w^{k}\propto e^{-\left(F(X)-F_{0}(X)\right)},\nonumber
\end{equation}
where the terms involving the $\phi$'s and the Jacobian (see (\ref{eq:weightsGeneral})) need not show up because they are constant for all particles, and thus drop out once the weights are scaled so that their sum equals one. This map is very efficient for this problem, because $L$ is a easy to compute (and can be computed offline). In particular, the evaluation of~(\ref{eq:SampleLorenz}) takes about $0.6$\% of the time it takes to carry out the variational minimization so that the cost of sampling is small compared to the cost of minimizing $F$. In this example, turning the 4D-Var code of the previous subsection into implicit particle smoothing code comes at an acceptable additional cost.

\subsubsection{Numerical results}
\begin{figure}[htbp]
\begin{center}
{\includegraphics[width=0.9\textwidth]{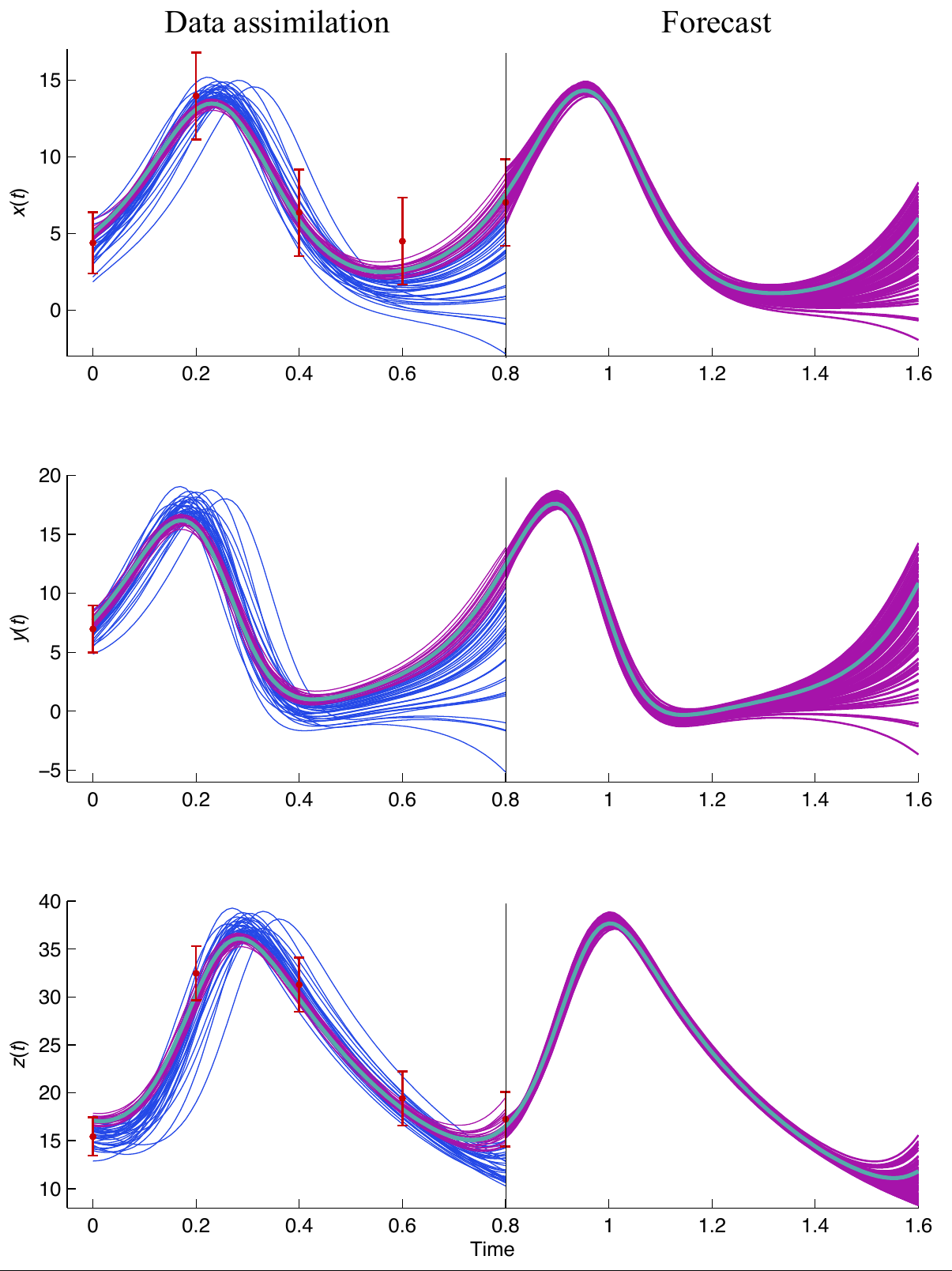}}
\caption{Illustration of data assimilation and forecasting using the implicit particle smoother. On the left (Time $\leq0.8$): 30 samples from the prior pdf (blue lines); the data and error bars (red); 25 samples obtained by the implicit particle smoother (purple); and the true state trajectory (teal). On the right (Time $>0.8$): 50 samples of a Gaussian approximation of the pdf of the state at time $0.8$ obtained by the implicit particle smoother (purple); and the true state trajectory (teal).} 
\label{fig:StrongConstraintFigure} 
\end{center}
\end{figure}

Figure \ref{fig:StrongConstraintFigure} illustrates the data assimilation with the implicit particle smoother. On the left (time $t\leq0.8$), we show the true state trajectory (teal), which was obtained by integrating the equations (\ref{eq:Lorenz}) starting from an initial condition which we got by sampling the prior pdf $p_0$. We also show the data (red dots) with error bars that represent two standard deviations ($2\sqrt{2}$ in our case) and the mean (red dot at time $0$) of the prior pdf with the same error bars. The blue lines show $30$ samples from the prior pdf and the purple lines are $25$ samples we obtained using the implicit particle smoother.
The sample mean (obtained by using 100 particles) is not shown, but it is very close to the true state trajectory. We can observe in this figure that the implicit particle smoother generates samples within the high probability region, because all samples are compatible with the data (they are within $2$ standard deviation of the data).

We can use the implicit particle smoother to make and assess a forecast (for time $t\geq 0.8$) as follows. We can approximate the pdf of the state at time $0.8$ by a Gaussian whose mean and covariance matrix can be computed from the weighted samples. We can then integrate samples, say 50, from this Gaussian. The result is shown as purple lines on the right of figure \ref{fig:StrongConstraintFigure}, and we observe that the true state (teal) is well within the cloud of samples. We can also observe that the uncertainty grows dramatically for times larger that $1.4$, i.e. a forecast should not be expected to be very accurate.

We further assessed the accuracy and reliability of the implicit particle smoother by running 100 twin experiments. A twin experiment amounts to generating a ``true'' initial condition by sampling the prior pdf $p_0$, integrating this initial condition forward in time and collecting observations by perturbing the true state trajectory with appropriate noise. The data are passed to the implicit particle smoother, which then produces an approximation to the conditional mean, which in turn is the minimum mean square error estimate of the initial condition. We than compute the Euclidean norm of the difference between the true initial condition and its approximation by the implicit particle smoother. The mean and standard deviation of this error norm, scaled by the mean of the norm of the true initial conditions, indicate the errors one should expect in each run. 

We compare the implicit particle smoother to the variational data assimilation scheme (4D-Var) which is implemented as part of the implicit particle smoother. In order to check that our implementation of the implicit particle smoother is free of errors, we compare its errors to those obtained with a Bayesian bootstrap method \cite{Doucet2001}. The Bayesian bootstrap method is an importance sampling method that uses the prior pdf $p_0$ as the importance function, i.e. we obtain samples from the prior pdf and then assign a weight based on the observations to each sample. The conditional mean can be approximated by the weighted sample mean and, for a large number of particles, this method converges to the true conditional mean. We observed that this method has converged for 1000 particles. With this number of particles, the Bayesian bootstrap method is about $4$ times slower than our variational implementation of the implicit particle smoother. The results of 100 twin experiments are shown in table \ref{table:StrongConstraintErrors}
\begin{table}[htdp]
\begin{center}
\begin{tabular}{c c c}
&  &\\
4D-Var & Implicit particle smoother  &  Bayesian bootstrap\\
\hline
            & (100 particles)  			 &  (1000 particles)\\
0.060 / 0.025 & 0.043 / 0.018  &  0.042 / 0.017\\
\end{tabular}

\caption{Errors (mean / standard deviation) in the reconstruction of the true initial condition for three data assimilation techniques.}
\label{table:StrongConstraintErrors}
\end{center}
\end{table}

We observe that all three data assimilation methods do their job, since all three yield a small error and a small error variance. The Bayesian bootstrap method and the implicit particle smoother converge to the same errors (both approximate the conditional mean), however, it is clear that the implicit particle smoother is more efficient since it is 4 times faster than the Bayesian bootstrap method. The implicit particle smoother improved the estimate of the variational method through sampling, i.e. by computing the conditional mean instead of the conditional mode, at an acceptable additional computational cost. Moreover, the implicit particle smoother delivers a quantitative measure of the uncertainty of its state estimate, which can be used to propagate the uncertainty forward in time and to assess the uncertainty of forecasts (see figure~\ref{fig:StrongConstraintFigure}). 

We conclude that the implicit particle smoother is efficient and reliable in it its variational implementation and that the additional computational cost, when compared to a variational method, is acceptable in view of its clear advantages (smaller errors and state estimates that are equipped with measures of their uncertainty).

\subsection{The weak constraint problem}
We now consider a weak constraint problem and use a stochastic version of the Lorenz attractor
\begin{align}
\frac{dx^1}{dt}& = \sigma(x^2-x^1)+gdW^1,\nonumber \\
\frac{dx^2}{dt}& = x_1(\rho-x^3)-x^2+gdW^2\nonumber \\
\frac{dx^3}{dt}& = x_1 x^2-\beta x^3 +gdW^3,\nonumber
\end{align}
where $W^1,W^2$ and $W^3$ are independent Brownian motions and where $\sigma,\rho$ and $\beta$ are as in section \ref{sec:StrongProblemExample} and $g=1/\sqrt{2}$. We discretize these stochastic differential equations (SDE) using the Euler-Maruyama scheme with constant time step $\delta=10^{-3}$, so that the discretization is a good approximation to the continuous equations \cite{Kloeden}. With this choice the function $R_{j}(x_j)$ for the discrete recurrence (\ref{eq:DiscreteModel}) becomes
\begin{equation}
R(x_j)=x_{j}+f(x_{j})\delta, \nonumber
\end{equation}
where $x_j=\left(x^1_j,x^3_j,x^3_j\right)^T$ and 
$$
f(x_{j}) =\left(\sigma(x^2_j-x^1_j),x^1_j(\rho-x^3_j)-x^2_j, x^1_j x^2_j-\beta x^3_j\right)^T,
$$ 
and $Z_k\sim\mathcal{N}(0,\delta/2)$.

The observations are all three state variables, collected at times $t_{n_k}=k\cdot r\cdot \delta$, perturbed by Gaussian noise with mean zero and covariance matrix $Q=2I_3$. The data assimilation problem is particularly hard when the time between observations is greater than the characteristic time scale at which transitions are made between the two attractors, which, for our choice of parameters is about $T=1/2$ \cite{MillerTellus}. We consider two cases: (\emph{a}) $r= 400$, i.e. the gap between observations is $0.4$ dimensionless time units and smaller than the characteristic time scale $0.5$; and (\emph{b}) $r=800$, i.e. the gap between observations is $0.4$ dimensionless time units and larger than the characteristic time scale. In both cases we assimilate the data sequentially using the implicit particle filter of section \ref{sec:filter}.

\subsubsection{Variational implementation of the implicit particle filter}
The main computational challenge of the implicit particle filter is to find the minima of the $F^k$'s in (\ref{eq:FFilter}). We explained in section (\ref{sec:var comp}) that these $F^k$'s are related to the weak constraint 4D-Var cost function and that 4D-Var codes can be used to carry out the required minimizations. The various weak 4D-Var codes differ mainly in the extent to which approximate techniques, such as linearizations or Gaussian assumptions, are used. We decided not to favor any particular approximate version of weak constraint 4D-Var and, for that reason, computed the first and second derivatives of $F^k$ analytically and used a trust-region method for the minimizations (see e.g. \cite{Conn}). This corresponds to an ``ideal'' implementation of weak constraint 4D-Var, for which the control variable is the full state trajectory \cite{tremolet}.

The trust-region approach requires a Cholesky decomposition of the Hessian of $F^k$ at each iteration of the minimization algorithm. Since this Hessian is banded (with band width $6$), the cost of one iteration is $O(3m)$, where $m$ is the number of model steps between observations. The number of model steps between observations increases quickly as the (non-dimensional) time between observations increases, because we chose a small time step $\delta$ to ensure accuracy of the discretization of the SDE's. Because the cost of each iteration is relatively large for large gaps between observations, it is worthwhile to invest into generating ``good seeds'' to initialize the trust-region iteration, so that it converges quickly. 

We generated a seed as follows: for each time window between observations, we first obtain $\overline{x}_m=x_{n_{m}+1:n_{m+1}}$ by integrating the stochastic differential equation. We then calculate the ``residual vector'' $r=x_{n_{m+1}}-y_{m+1}$ and perturb the model path using $\overline{x}^{r}_{j}=\overline{x}_{j}-r(j/r)$ for each $j=0,1,2\dots,r$. This procedure rotates the model path $\overline{x}_j$ towards the observation. 

We refine this seed with a multi-grid technique, which is conceptually similar to the multi-grid finite difference method \cite{Fedorenko} and multi-grid Monte Carlo \cite{GoodmanSokal} (see also \cite{chains}). The idea is to first perform a cheap minimization on a coarse grid, i.e. with a larger time step, and then use the result of this minimization, interpolated onto the fine grid, as the seed for the minimization on the finer grid. The reason why we can use this multi-grid approach here is that the conditional pdf depends on the model (it is proportional to the product of the pdf for the model and the pdf for the observations), which in turn represents an approximation to an SDE. The conditional pdf we obtain with a model and time step say $\hat \delta<\delta$ should thus be somewhat similar to the conditional pdf we obtain with a time step $\delta <\hat \delta$. Since $F^k$ is minus the logarithm of the conditional density, we expect that the minimizer of $F^k$ with a model with time step $\tilde{\delta}$ is similar to the minimizer of an $F^k$ with a model and time step $\delta<\tilde{\delta}$. 

In addition to speeding up the minimization, the multi-grid approach proved effective to identify local minima of $F^k$. We observed in our experiments that the global minimum of $F^k$ was rarely larger than $10$, independent of the time step or even the gap between observation times. Local minima were observed to be as large as $200$. This observation can be used to identify local minima of $F^k$: the result of coarse grid minimization is rejected if the minimum is above a threshold $\phi_{c}$, and we restart the minimization with a new (unrefined) seed $\overline{x}_m$.

To test our minimization algorithm (the weak 4D-Var code), we compare its output to the output of a trust-region method that uses ``the truth" as its seed, i.e. we generate a reference state trajectory by integrating the SDE's, collect observations from this state trajectory and run our 4D-Var code as well as a trust-region method that is initialized with the true state trajectory. This should give us an idea of how accurate our 4D-Var code is, because the true state trajectory typically lies only a few Newton steps away from a relevant mode of the conditional pdf. We find that our multi-grid scheme finds the same minimum as seeding the minimization with the truth 100\% of the time for gaps between observations that are less than $1.5$ dimensionless time units ($1500$ model steps).

\subsubsection{Implementation of the map $\psi$}
\label{sec:WeakProblemSampling}
Upon minimization of the $F^k$'s, we solve (\ref{eq:FImplicitFilter}) for each particle to obtain samples from the conditional pdf. To solve this underdetermined equation, we use the same approach as in section \ref{sec:StrongProblemExample}, i.e. we replace $F^k$ in (\ref{eq:FImplicitFilter}) by its quadratic approximation and solve a quadratic equation. This approach is very efficient for this problem, because we can solve the quadratic equation using the Cholesky factor, $L$, of the Hessian of $F^k$, which is available from the trust-region minimization (the variational part of the implicit particle filter). The Jacobian of this map is easily calculated to be the determinant of $L$ (the product of its diagonal entries). Generating a sample using this map takes about 1/10000 of the time it takes to carry out the minimization. The cost of sampling is thus small compared to the cost of minimizing $F^k$, i.e. turning a weak 4D-Var code into implicit sampling code comes at a low additional cost. Again, we account for the error we make by replacing $F^k$ by its quadratic approximation through the weights, which become
\begin{equation}
\hat{w}^{k}=w^{k}e^{-\phi^k}\, e^{-\left(F(X^k_{n_m:n_{m+1}})-F_{0}(X^k_{n_m:n_{m+1}})\right)}\det L^{-1}.\nonumber
\end{equation}
Note that the factors with $\phi^k$ and the Jacobian of the map ($\det L^{-1}$) must appear in the weights because the functions $F^k$ are different for each particle and, thus, can have different minima and different Hessians.

\subsubsection{Monte Carlo variance reduction}
\label{sec:VarianceReduction}
We can improve the performance of the implicit particle filter by using standard Monte Carlo variance reduction techniques such as prior boosting, rejection control or partial rejection control \cite{GordonSIR,LiuinDoucet,rejcon}. These methods rely on generating an expanded ensemble of particles from which only a subset will be promoted to the next assimilation window. It is important to realize that the expanded ensemble of particles does not require additional minimizations, because the new ``intermediate'' particles share their $F^k$'s with their ``parent'' particles (for which the minimization has already been carried out).

In particular, we can generate $m>1$ ``intermediate'' particles for each of the $M$ particles by using (\ref{eq:SampleLorenz}) repeatedly. We thus obtain $mM$ samples of the conditional pdf, essentially at the cost of $M$ samples (since using~(\ref{eq:SampleLorenz}) is cheap compared to the minimization of $F^k$). This ``prior boosting'' technique proved effective at increasing sample diversity in our numerical experiments.

\subsubsection{Numerical results}
We test the efficiency and accuracy of the implicit particle filter by running twin experiments, as we did in section \ref{sec:StrongProblemExample}. Each twin experiment amounts to generating a reference solution up to time $4$, also called ``the truth," using the Euler-Maruyama discretization of the stochastic Lorenz attractor, and collecting observations at times $t_{n_{k}}=k\cdot r\cdot \delta$. We consider two cases: (\emph{a}) data is collected every $r=400$ model steps (the gap between observations is smaller than the characteristic time scale of the Lorenz attractor); and (\emph{b})  data is collected every $r=800$ model steps (the gap between observations is larger than the characteristic time scale of the Lorenz attractor). In each case, the data are passed to three data assimilation algorithms: (\emph{i}) the implicit particle filter in its sequential form (see section \ref{sec:filter}); (\emph{ii}) the Bayesian bootstrap filter with resampling (also sometimes known as the standard SIR filter), which uses the pdf $p(x_{n_m+1:n_{m+1}}|x_{n_m})$ as its importance function \cite{GordonSIR,Doucet2001}; and~(\emph{iii}) an implementation of weak constraint 4D-Var, which uses the same (nonlinear) multi-grid trust-region method as the implicit particle filter to carry out the minimizations. The weak 4D-Var code also assimilates the observations sequentially. The output of the two filters is an approximation of the conditional mean, and the output of the weak constraint 4D-Var code is an approximation of the conditional mode. 

\begin{figure}[h!]
\begin{center}
{\includegraphics[width=1.0\textwidth]{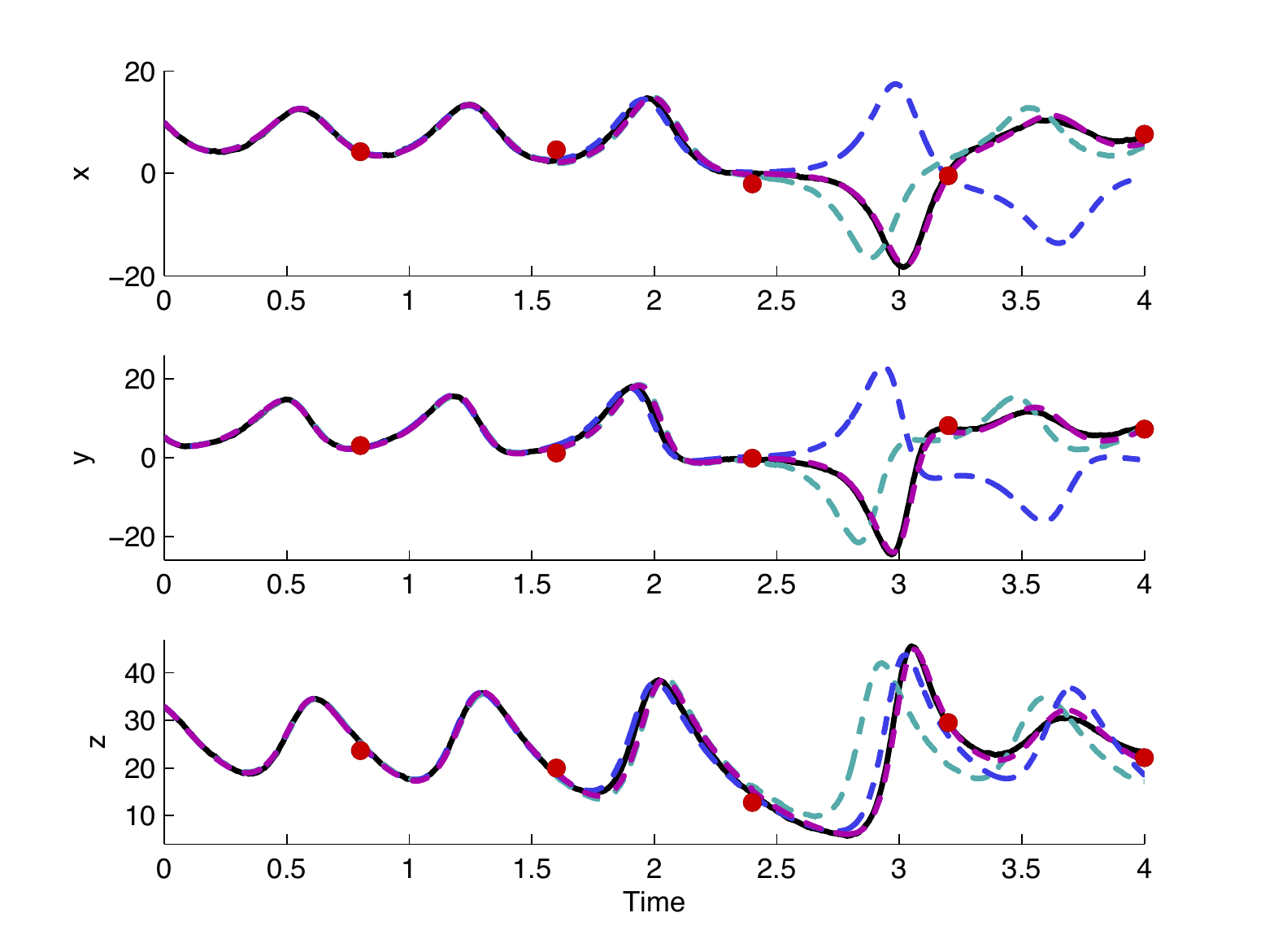}}
\caption{Reconstructions of a reference path (solid-black) from a set of 5 observations (red dots) by three data assimilation methods. Dashed-teal: reconstruction by the standard SIR filter with 20 particles. Dashed-blue: reconstruction by weak constraint 4D-Var. Dashed-purple: reconstruction by the implicit particle filter with 10 particles, each with 50 intermediate particles.} 
\label{fig:eyeball} 
\end{center}
\end{figure}

In figure \ref{fig:eyeball} we plot the results of one twin experiment, where we assimilate sequentially 5 observations, with $r=800$ model steps (0.8 dimensionless time units) between observations (case (\emph{b})).
We observe that, with $20$ particles, the SIR filter looses track of the true state trajectory after a relatively short time. The reason is that none of the samples is sufficiently close to the observations, i.e. we observe the typical effect of sample impoverishment. The weak constraint 4D-Var code can not follow the true state trajectory, because, starting at time $2.4$, it is trapped in a local minimum. The implicit particle filter with 20 particles, each boosted with 50 intermediate particles (see section \ref{sec:VarianceReduction}) can follow the true state trajectory at all times. The reason why the implicit particle filter is not ``stuck'' in a local minimum (as is 4D-Var) is that it is able to track the various modes of the conditional pdf, since the minimization is performed particle by particle. In this example, about 10 particles appear sufficient to track all relevant modes. 

We perform 100 such twin experiments, because a single twin experiment is not very informative (it is a random event).  For each one we compute the errors $e=x^{F}_{0:n}-x_{0:n}$, where $x_{0:N}$ is the true state trajectory and $x^{F}_{0:N}$ is the output of the data assimilation (implicit particle filter, SIR filter, or 4D-Var). The mean and standard deviation of the Euclidean norm of these errors indicates the errors one can expect for each method and in each run. The results are shown in table \ref{table:error}, where we scaled the errors and their standard deviations by the mean of the Euclidean norm of the true state trajectory.
\begin{table}[htdp]
\begin{center}
\begin{tabular}{c c c c }
\multicolumn{4}{c}{Case (\emph{a}): $r=400$ model steps between observations}\\
\hline
  Number of particles 		& 4D-Var  		   &  IPF 			 & SIR\\
 -		& 0.086 / 0.063 & -      		 & - \\
10 		& - 			   & 0.042 / 0.012 &  0.15 / 0.16 \\
20 		& - 		        & 0.040 / 0.013 & 0.092 / 0.10 \\
100 	& - 		        & - 			&0.048 / 0.050 \\
1000 	& - 		        &- 			&0.038 / 0.013 \\
5000 	& - 			  & - 			& 0.037 / 0.037 \\
\multicolumn{4}{c}{ }\\
\multicolumn{4}{c}{Case (\emph{b}): $r=800$ model steps between observations}\\
\hline
Number of particles  		& 4D-Var  		   &  IPF 			 & SIR\\
 -		& 0.13 / 0.15 & -    & -\\
10 		& -	     & 0.074 / 0.070    &  0.18 / 0.17 \\
20 		& -	     & 0.074 / 0.080    & 0.14 / 0.15 \\
100 		& - 	     & - 			& 0.077 / 0.082 \\
1000 	& -	     & -			& 0.065 / 0.056 \\
5000 	& - 	    &-			& 0.064 / 0.064 \\
&\multicolumn{3}{c}{}\\
\end{tabular}
\end{center}
\caption{Errors (mean / standard deviation) of three data assimilation techniques. 4D-Var: an ideal implementation of weak constraint 4D-Var. IPF: the implicit particle filter (each particle has $50$ intermediate particles). SIR: the Bayesian bootstrap filter.}
\label{table:error}
\end{table}

We observe from table \ref{table:error}, that the implicit particle filter as well as the standard SIR filters can provide accurate approximations of the true state in both cases (since all errors are relatively small), provided that the number if particles is large enough. What is important to realize here is that the implicit particle filter can achieve a similar accuracy, but with a significantly lower number of particles than the standard SIR filter. The weak 4D-Var method can not achieve the accuracy of the particle filters, especially if the gap between observations is larger (case (\emph{b})). The reason is  that the method is trapped in local minima, i.e. 4D-Var is unable to track more than one mode. The implicit particle filter on the other hand is able to track all relevant modes (at least in this example), due to the particle-by-particle minimization. The cost is that the implicit particle filter performs a 4D-Var calculation for each particle, i.e. the computational cost is roughly the cost of the 4D-Var method times the number of particles, because sampling is cheap compared to the minimization. The pay-off of being able to track all relevant modes however outweighs this cost in this example. Moreover, the implicit particle filter is very easy to parallelize (the particles only communicate in the resampling step), and a parallel implementation of the implicit particle filter is comparable in computation time to a (serial) implementation of weak 4D-Var.

To further assess the quality of the implicit particle filter, we compute the normalized effective sample size
\begin{equation}
\frac{ M_{\mbox{eff}}}{M}=\frac{(\sum_{k=1}^{M}w_{k})^2}{M\sum_{k=1}^{M}w_{k}^{2}},\nonumber
\end{equation}
where $M$ is the number of particles, for each twin experiment at the last data assimilation cycle. The normalized effective sample size indicates the percentage of particles that contribute meaningfully to the approximation of the conditional pdf \cite{Doucet2001} and we compare the normalized effective sample size of the implicit particle filter and the SIR filter. The results are shown in table~\ref{table:EffectiveSampleSize}.
\begin{table}[htdp]
\begin{center}
\begin{tabular}{c c c}
\multicolumn{3}{c}{Case (\emph{a}): $r=400$ model steps between observations}\\
\hline
 Number of particles  &  IPF 			 & SIR\\
10 		& 95.0\% & 50.4 \% \\
20 		& 94.5 \% & 49.2 \% \\
100  & - & 49.0 \%\\
1000 & - & 48.5\% \\	
5000 & - & 48.6\%\\	
\multicolumn{3}{c}{}\\
\multicolumn{3}{c}{Case (\emph{b}): $r=800$ model steps between observations}\\
\hline
 Number of particles  &  IPF 			 & SIR\\
10 		& 84.8\% & 37.9 \% \\
20 		& 84.1 \% & 34.7 \% \\
100  & - & 32.6 \%\\
1000 & - & 33.0\% \\	
5000 & - & 33.0\%\\
\end{tabular}
\end{center}
\caption{Normalized effective sample size of the implicit particle filter and the standard SIR filter}
\label{table:EffectiveSampleSize}
\end{table}

We observe that, with a relatively short time between observations (case (\emph{a})), about 50\% of the particles of the standard SIR filter are contributing meaningfully to the ensemble averages. The situation is more dramatic for a larger gap between observations (case (\emph{b})), where we observe effective sample sizes of about 35\%. The normalized effective sample size of the implicit particle filter is about $95$\% for small gaps, and about $84$\% for larger gaps. In summary, we conclude that the implicit particle filter performs accurately and reliably on our test problems and yields accurate results (with uncertainty quantifications) at a reasonable computational cost.

\section{Conclusions}
\label{sec:Conclusions}
The implicit particle filter was introduced in \cite{chorintu2010,chorintupnas,Morzfeld2011} as a sequential Monte Carlo method for data assimilation. In the present paper, we derived the implicit particle filter in a more general set up and presented extensions to implicit particle smoothing and to data assimilation for perfect models.

We explored the connection of these implicit particle methods with variational data assimilation and showed that existing variational codes can be used for efficient implementation of implicit particle methods. In particular, we showed that variational codes can carry out the minimizations required by implicit particle methods. Turning a variational code into an implicit particle method then amounts to solving an underdetermined scalar equation; methods to solve these equations efficiently can be found in our earlier work (e.g. in \cite{chorintu2010,Morzfeld2011}). The additional cost of implicit particle methods is thus small, and the payoff is that one can obtain the minimum mean square error estimate of the state along with a quantitative measure of its uncertainty, whereas variational codes produce biased state estimates with at best approximate error quantifications.

We have demonstrated the applicability and efficiency of the implicit particle methods by applying them to the Lorenz  attractor. We considered the strong constraint data assimilation problem (estimation of initial conditions for a perfect model) as well as the weak constraint problem (estimation of the state trajectory of an uncertain model) and, in both cases discussed the details of the variational aspects of the filter. In the strong constraint problem, we found that the implicit particle filter can improve the variational estimate significantly by turning the conditional mode into the conditional mean (the minimum mean square error estimator). Moreover, the implicit particle smoother produced quantitative measures of the uncertainty which were useful in assessing the uncertainty in forecasts. In the weak constraint problem, we found that the implicit particle filter requires about 1\% of the particles of a standard SIR filter, and that it performs better than weak constraint 4D-Var because it can track all relevant modes of the conditional pdf. In every case we considered, the cost of solving the implicit equations to generate samples was small compared to the cost of the minimizations, i.e. to the cost the implicit particle filter shares with variational data assimilation.

\section*{Acknowledgements}
We would like to thank our collaborators at Oregon State University, Professors Robert Miller and Yvette Spitz, and Dr. Brad Weir, for helpful discussion and comments. This work was supported in part by the Director, Office of Science, Computational and Technology Research, U.S. Department of Energy under Contract No. DE-AC02-05CH11231, and by the National Science Foundation under grants DMS-0705910 and OCE-0934298.

\bibliographystyle{plain}
\bibliography{implicit}

\begin{thebibliography}{10}

\bibitem{Bennet1993}
A.~F. Bennet, L.~M. Leslie, C.~R. Hagelberg, and P.~E. Powers.
\newblock A cyclone prediction using a barotropic model initialized by a
  general inverse method.
\newblock {\em Monthly Weather Review}, 121:1714--1728, 1993.

\bibitem{BickelBootstrap}
P.~Bickel, T.~Bengtsson, and J.~Anderson.
\newblock Sharp failure rates for the bootstrap particle filter in high
  dimensions.
\newblock {\em Pushing the Limits of Contemporary Statistics: Contributions in
  Honor of Jayanta K. Ghosh}, 3:318--329, 2008.

\bibitem{Carpenter}
J.~Carpenter, P.~Clifford, and P.~Fearnhead.
\newblock Improved particle filter for nonlinear problems.
\newblock {\em Radar, Sonar and Navigation, IEE Proceedings -}, 146(1):2 --7,
  feb 1999.

\bibitem{chains}
A.~Chorin.
\newblock Monte carlo without chains.
\newblock {\em Communications in Applied Mathematics and Computational
  Science}, 3:77--93, 2008.

\bibitem{ChorinHald}
A.~J. Chorin and O.~H. Hald.
\newblock {\em Stochastic Tools in Mathematics and Science}.
\newblock Springer, first edition, 2006.

\bibitem{Chorin2004}
A.~J. Chorin and P.~Krause.
\newblock Dimensional reduction for a {B}ayesian filter.
\newblock {\em Proceedings of the National Academy of Sciences},
  101(42):15013--15017, 2004.

\bibitem{chorintu2010}
A.~J. Chorin, M.~Morzfeld, and X.~Tu.
\newblock Implicit particle filters for data assimilation.
\newblock {\em Communications in Applied Mathematics and Computational
  Science}, 5(2):221--240, 2010.

\bibitem{chorintupnas}
A.~J. Chorin and X.~Tu.
\newblock Implicit sampling for particle filters.
\newblock {\em Proceedings of the National Academy of Sciences},
  106(41):17249--17254, 2009.

\bibitem{Courtier1997}
P.~Courtier.
\newblock Dual formulation of four-dimensional variational data assimilation.
\newblock {\em Quarterly Journal of the Royal Meteorological Society},
  123:2449--2461, 1997.

\bibitem{Courtier1994}
P.~Courtier, J.N. Thepaut, and A.~Hollingsworth.
\newblock A strategy for operational implementation of 4d-var, using an
  incremental approach.
\newblock {\em Quarterly Journal of the Royal Meteorological Society},
  120:1367--1387, 1994.

\bibitem{DimetTalagrand}
F.~X.~Le Dimet and O.~Talagrand.
\newblock Variational algorithms for analysis and assimilation of
  meteorological observations: theoretical aspects.
\newblock {\em Tellus A}, 38A(2):97--110, 1986.

\bibitem{Doucet2001}
A.~Doucet, N.~de~Freitas, and N.~Gordon, editors.
\newblock {\em Sequential {M}onte {C}arlo {M}ethods in practice}.
\newblock Springer, 2001.

\bibitem{Doucet}
A.~Doucet, S.~Godsill, and C.~Andrieu.
\newblock On sequential monte carlo sampling methods for bayesian filtering.
\newblock {\em Statistics and Computing}, 10:197--208, 2000.

\bibitem{EvensenEnKF}
G.~Evensen.
\newblock Sequential data assimilation with a nonlinear quasi-geostrophic model
  using monte carlo methods to forecast error statistics.
\newblock {\em Journal of Geophysical Research}, 99:10143--10162, 1994.

\bibitem{EvensenLorenz}
G.~Evensen.
\newblock Advanced data assimilation for strongly nonlinear dynamics.
\newblock {\em Monthly Weather Review}, 125:1342--1354, 1997.

\bibitem{EvensenBook}
G.~Evensen.
\newblock {\em Data assimilation: the ensemble Kalman filter}.
\newblock Springer, 2006.

\bibitem{Fedorenko}
R.~P. Fedorenko.
\newblock A relaxation method for solving elliptic difference equations.
\newblock {\em USSR Computational Mathematics and Mathematical Physics}, 1,
  1961.

\bibitem{Fletcher}
R.~Fletcher.
\newblock {\em Practical Methods of Optimization}.
\newblock Wiley, second edition, 1987.

\bibitem{Fournier2010}
A.~Fournier, G.~Hulot, D.~Jault, W.~Kuang, A.~Tangborn, N.~Gillet, E.~Canet,
  J.~Aubert, and F.~Lhuillier.
\newblock An introduction to data assimilation and predictability in
  geomagnetism.
\newblock {\em Space Science Reviews}, 155(1-4):247--291, 2010.

\bibitem{Geweke}
J.~Geweke.
\newblock Bayesian inference in econometric models using monte carlo
  integration.
\newblock {\em Econometrica}, 24:1317 -- 1399, 1989.

\bibitem{GoodmanSokal}
J.~Goodman and A.~D. Sokal.
\newblock Multigrid monte carlo method. conceptual foundations.
\newblock {\em Phys. Rev. D}, 40:2035--2071, Sep 1989.

\bibitem{GordonSIR}
N.~J. Gordon, D.~J. Salmond, and A.~F.~M. Smith.
\newblock Novel approach to nonlinear/non-gaussian bayesian state estimation.
\newblock {\em Radar and Signal Processing, IEE Proceedings F}, 140(2):107
  --113, apr 1993.

\bibitem{Hammersley}
J.~M. Hammersley and D.C. Handscomb.
\newblock {\em Monte Carlo Methods}, volume~1.
\newblock Methuen young books, 1 edition, 1964.

\bibitem{JohansenDoucet}
A.~M. Johansen and A.~Doucet.
\newblock A note on auxiliary particle filters.
\newblock {\em Statistics \& Probability Letters}, 78(12):1498 -- 1504, 2008.

\bibitem{Kalman1960}
R.~E. Kalman.
\newblock A new approach to linear filtering and prediction theory.
\newblock {\em Transactions of the ASME--Journal of Basic Engineering},
  82(Series D):35--48, 1960.

\bibitem{Kalman1961}
R.~E. Kalman and R.~S. Bucy.
\newblock New results in linear filtering and prediction theory.
\newblock {\em Transactions of the ASME--Journal of Basic Engineering},
  83(Series D):95--108, 1961.

\bibitem{Kalnay}
E.~Kalnay.
\newblock {\em Atmospheric modeling, data assimilation and predictabilty}.
\newblock Cambridge University Press, 2003.

\bibitem{KalnayEnkFvs4DVar}
E.~Kalnay, H.~Li, T.~Miyoshi, S.~C. Yang, and J.~Ballabrera-Poy.
\newblock 4-d-var or ensemble kalman filter.
\newblock {\em Tellusl}, 59A:758--773, 2007.

\bibitem{KalosWhitlock}
M.~H. Kalos and P.~A. Whitlock.
\newblock {\em Monte Carlo Methods}, volume~1.
\newblock John Wiley \& Sons, 1 edition, 1986.

\bibitem{Kloeden}
P.~E. Kloeden and E.~Platen.
\newblock {\em Numerical Solution of Stochastic Differential Equations}.
\newblock Springer, 3 edition, 1999.

\bibitem{Kurapov2007}
A.~Kurapov, G.~D. Egbert, J.~S. Allen, and R.~N. Miller.
\newblock Representer- based variational data assimilation in a nonlinear model
  of nearshore circulation.
\newblock {\em Journal of Geophysical Research}, 112:C11019, 2007.

\bibitem{LiuChen1995}
J.~S. Liu and R.~Chen.
\newblock Blind deconvolution via sequential imputations.
\newblock {\em Journal of the American Statistical Association}, 90(430):pp.
  567--576, 1995.

\bibitem{LiuinDoucet}
J.~S. Liu, R.~Chen, and T.~Logvinenko.
\newblock A theoretical framework for sequential importance sampling with
  resampling.
\newblock In A.~Doucet, N.~de~Freitas, and N.~Gordon, editors, {\em Sequential
  {M}onte {C}arlo {M}ethods in practice}, chapter~11. Springer, 2001.

\bibitem{rejcon}
J.~S. Liu, R.~Chen, and W.~H. Wong.
\newblock Rejection control and sequential importance sampling.
\newblock {\em Journal of the American Statistical Association}, 93(443):pp.
  1022--1031, 1998.

\bibitem{Lorenz63}
E.~N. Lorenz.
\newblock {Deterministic Nonperiodic Flow.}
\newblock {\em Journal of Atmospheric Sciences}, 20:130--148, March 1963.

\bibitem{MillerTellus}
R.~N. Miller, E.~F. Carter, and S.~T. Blue.
\newblock Data assimilation into nonlinear stochastic models.
\newblock {\em Tellus A}, 51(2):167--194, 1999.

\bibitem{DelMoral2012}
P.~Del Moral, A.~Doucet, and A.~Jasra.
\newblock On adaptive resampling strategies for sequential monte carlo methods.
\newblock {\em Bernoulli}, 18(1):252--278, 2012.

\bibitem{Morzfeld2012}
M.~Morzfeld and A.~J. Chorin.
\newblock Implicit particle filtering for models with partial noise, and an
  application to geomagnetic data assimilation.
\newblock {\em Nonlinear Processes in Geophysics}, page submitted for
  publication, 2012.

\bibitem{Morzfeld2011}
M.~Morzfeld, X.~Tu, E.~Atkins, and A.~J. Chorin.
\newblock A random map implementation of implicit filters.
\newblock {\em Journal of Computational Physics}, 231(0):--, 2011.

\bibitem{Nocedal}
J.~Nocedal and S.~T. Wright.
\newblock {\em Numerical Optimization}.
\newblock Springer, second edition, 2006.

\bibitem{Rabier4DVAR}
F.~Rabier and P.~Courtier.
\newblock Four-dimensional assimilation in the presence of baroclinic
  instability.
\newblock {\em Quarterly Journal of the Royal Meteorological Society},
  118(506):649--672, 1992.

\bibitem{Conn}
A.~R.Conn, N.~I.~M. Gould, and P.~L. Toint.
\newblock {\em Trust-region methods}.
\newblock Society for Industrial and Applied Mathematics, first edition, 2000.

\bibitem{GordonReview}
M.~S.Arulampalam, S.~Maskell, N.~Gordon, and T.~Clapp.
\newblock A tutorial on particle filters for online nonlinear/non-gaussian
  bayesian tracking.
\newblock {\em Signal Processing, IEEE Transactions on}, 50(2):174 --188, feb
  2002.

\bibitem{GelfandSmithResamp}
A.~F.~M. Smith and A.~E. Gelfand.
\newblock Bayesian statistics without tears: A sampling-resampling perspective.
\newblock {\em The American Statistician}, 46(2):pp. 84--88, 1992.

\bibitem{Bickel}
C.~Snyder, T.~Bengtsson, P.~Bickel, and J.~Anderson.
\newblock Obstacles to high-dimensional particle filtering.
\newblock {\em Monthly Weather Review}, 136(12):4629--4640, Dec 2008.

\bibitem{Talagrand1997}
O.~Talagrand.
\newblock Assimilation of observations, an introduction.
\newblock {\em Journal of the Meteorological Society of Japan}, 75(1):191--209,
  1997.

\bibitem{TalagrandCourtier}
O.~Talagrand and P.~Courtier.
\newblock Variational assimilation of meteorological observations with the
  adjoint vorticity equation. i: Theory.
\newblock {\em Quarterly Journal of the Royal Meteorological Society},
  113(478):1311--1328, 1987.

\bibitem{tremolet}
Y.~Tremolet.
\newblock Accounting for an imperfect model in 4d-var.
\newblock {\em Quarterly Journal of the Royal Meteorological Society},
  132(621):2483--2504, 2006.

\bibitem{vanLeeuwen}
P.~J. van Leeuwen.
\newblock Nonlinear data assimilation in geosciences: an extremely efficient
  particle filter.
\newblock {\em Quarterly Journal of the Royal Meteorological Society},
  136(653):1991--1999, 2010.

\bibitem{PeterJan2011}
P.~J. van Leeuwen.
\newblock Efficient nonlinear data-assimilation in geophysical fluid dynamics.
\newblock {\em Computers \& Fluids}, 46(1):52 -- 58, 2011.

\bibitem{Zupanski1997}
D.~Zupanski.
\newblock A general weak constraint applicable to operational 4{DVAR} data
  assimilation systems.
\newblock {\em Monthly Weather Review}, 125:2274--2292, 1997.

\end{thebibliography}
  \end{document}